\documentclass[pre,aps,preprint,showpacs]{revtex4-1}
\usepackage{revsymb4-1,latexsym,amssymb,amsmath,comment}
\usepackage{graphicx,psfrag,subfigure,stmaryrd}
\usepackage[ansinew]{inputenc}

\newcommand{\be}{\begin{equation}}
\newcommand{\ee}{\end{equation}}
\newcommand{\bel}[1]{\begin{equation}\label{#1}}
\newcommand{\bea}{\begin{eqnarray}}
\newcommand{\eea}{\end{eqnarray}}
\newcommand{\ba}{\begin{array}}
\newcommand{\ea}{\end{array}}
\newcommand{\bef}{\begin{figure}}
\newcommand{\ef}{\end{figure}}

\begin{document}

\author{Thomas Bose and Steffen Trimper}
\affiliation{Institute of Physics,
Martin-Luther-University, D-06099 Halle, Germany}
\email{thomas.bose@physik.uni-halle.de}
\email{steffen.trimper@physik.uni-halle.de}
\title{Noise-assisted tumor-immune cells interaction}
\date{\today }

\begin{abstract}
We consider a three-state model comprising tumor cells, effector cells and tumor detecting cells 
under the influence of noises. It is demonstrated that inevitable stochastic forces existing in all 
three cell species are able to suppress tumor cell growth completely. Whereas the deterministic model 
does not reveal a stable tumor-free state, the auto-correlated noise combined with cross-correlation functions  
can either lead to tumor dormant states, tumor progression as well as to an elimination of tumor cells. 
The auto-correlation function exhibits a finite correlation time $\tau$ while the cross-correlation functions 
shows a white noise behavior. The evolution of each of the three kinds of cells leads to a multiplicative noise coupling. 
The model is investigated by means of a multivariate Fokker-Planck equation for small $\tau$.  
The different behavior of the system is above all determined by the variation of the correlation time 
and the strength of the cross-correlation between tumor and tumor detecting cells. The theoretical model is 
based on a biological background discussed in detail and the results are tested using realistic parameters from 
experimental observations. 


\end{abstract}

\maketitle

\section{Introduction}

Tumor growth has become an important issue in medicine, biology and physics. The understanding of cancer 
growth mechanisms is necessary to develop relevant strategies against the disease. In the past, deterministic 
models have been proposed for interacting tumor and immune cells which are investigated by performing stability
and bifurcation analysis \cite{Kuznetsov:BullMathBiol:56:1994,KirschnerPanetta:JMathBiol:1998,Kuznetsov:MathCompMod33:2001:1275}. 
Moreover, a deterministic mathematical model with strong relation to experimental data is presented in 
\cite{DePillis:CancerRes:65:7950:2005}. As a new aspect the delay time between the detection of tumor cells 
by the immune system and the arrival of activated killer cells at the tumor site was
taken into account in \cite{Rodriguez-P:MathMedBiol:24:2007}. All these mathematical models can be considered 
as two state models of predator-prey-type. In general, such models can show interesting behavior as demonstrated 
in many examples in \cite{Murray:MathematicalBiology:1993}. Recently Ref. \cite{Khasin:PRE83:2011:031917} has discussed 
the effect of deterministically imposed transitions in reaction and population systems on the rates of rare 
events such as a crossing-over to population extinction. Another approach was chosen in 
\cite{BenAmar:PhysRevLett.106.148101} where the early stages of tumor growth was investigated. More precise, 
the geometrical aspect of contour instabilities was related to cell-cell interactions. Likewise the role of 
noisy influences can be regarded. As a result the stochastic forces may change the dynamics, in particular it 
was shown that the evolutionary dynamics is altered in case demographic noises are included in a 
deterministic model of interacting players \cite{BladonPhysRevE.81.066122}. As well, intrinsic stochasticity 
was considered in \cite{Parker:PhysRevE.80.021129} applied to the Lotka-Volterra model with special emphasis 
on the elimination of species. In addition, the extinction of stochastic populations caused by intrinsic noise 
was analyzed in \cite{Assaf:PhysRevE.81.021116}. Regarding tumor evolution one often refers to a logistic 
growth model which offers relevant results in spite of its simplicity \cite{BoseTrimper:PRE:2009:051903}. 
In the present paper we also use as the basic model the logistic equation for the deterministic cancer cell 
growth dynamics, see Eq.~\eqref{detmod1}. Recently a generalized logistic equation was studied by supplementing 
the birth rate by a Markovian dichotomic noise \cite{Aquino:EPL89:2010:50012}. Another essential point is that the 
tumor genesis is often accompanied by an abnormal proliferative activity of human tissue. In 
\cite{DiGarbo:PhysRevE.81.061909} the authors have reported on a mathematical model which covers the growth 
properties in terms of a variable renewal rate of cell populations in colon crypts. A further class of models 
is related to a single population where only the tumor cells are considered as the relevant variable. Here 
the deterministic equation are subjected to additional random forces which allows an analysis in terms of the 
related Fokker-Planck equation. Models for for white noise \cite{Ai:PRE:67:022903} and for colored noise 
\cite{BoseTrimper:PRE:2009:051903} has been predicted. The latter one contains tumor-immune interactions in an 
implicit manner. Later a modified model was investigated by introducing a bounded noise which mimics the 
reduction of the tumor size due to a possible immune response \cite{DOnofrio:PhysRevE.81.021923}. Therefore, 
the random nature is also attributed to the immune system.\\ 
This idea plays likewise a significant role in the present approach. Different to former works 
\cite{BoseTrimper:PRE:2009:051903,DOnofrio:PhysRevE.81.021923} the tumor-immune interplay is now incorporated 
explicitly. However the main point is that we demonstrate tumor-immune cell reactions can be induced by 
stochastic forces. To be more specific our model describes the time evolution of three different cell types: 
(i) tumor cells the density of which is denoted by $X(t)$, (ii) effector cells with density $Y(t)$ and 
(iii) tumor detecting cells with density $Z(t)$. Whereas the last kind of cells is only able to recognize 
tumor cells but not to kill them. The effector cells have the ability to eliminate tumor cells. The 
deterministic model introduced in Sec.~III describes the mutual interaction between the three species. 
However this model offers no stable tumor-free state. Due to the inclusion of inevitable randomness 
the growth and death rates of the immune and tumor cells, respectively, are altered
immediately. Toward a more realistic description we allow also the occurrence of cross-correlations between the 
noise acting on the tumor cells and that one acting on the detecting cells. The resulting set of stochastic 
equations with multiplicative noise can be transfered to the related Fokker-Planck equation. By variation of 
the strength of the cross correlation and the finite the correlation time the system tends to different stable states 
which differ from those of the deterministic system. Especially we show that the noisy system exhibit the complete 
suppression of the tumor. The paper is organized as follows. First, we present in Sec.~II some biological 
ideas concerning the tumor-immune interaction  The mathematical model is developed and discussed in Sec.~III. 
Due to the inclusion of randomness the related Fokker-Planck equation is introduced in Sec.~IV. Afterwards we 
present our results in Sec.~V before we finish with some conclusions in Sec.~VI.

\section{Biological motivation}

Before we present in the forthcoming section the mathematical model let us summarize 
some biological mechanisms concerning the interaction between the tumor and the immune system. 
In particular, this section is focused on the main underlying ideas which are necessary 
for the understanding of our presented model. Introduced in the early 1900's \cite{EhrlichPaul:1909} 
and again suggested in the middle of the 20th century \cite{Burnet:1957,Thomas:1959} there is 
the hypothesis that the immune system is able to detect and to eliminate nascent transformed cells. 
During the last decade the concept of the immune surveillance of the body played a significant role 
in tumor elimination, too. The investigations are supported by experimental results verifying the immune 
surveillance hypothesis \cite{Shankaran:Nature410:2001:1107}(and Refs. therein). Furthermore, the immune
surveillance concept was modified and is now known as 'immunoediting' which
reflects the dual role of the immune response during the early stages of cancer
growth \cite{Dunn:NatImmun3:2002:991,Kim:Immunology:121:2007:1,Schreiber:Science331:2011:1565}.
The term immunoediting means both the ability of the immune system to destroy the tumor cells 
and a possible sculpting of the cancer cells. As the result all cells with a low immunogenicity 
will survive and begin to proliferate. This escape of the tumor from the control of
the immune system can be regarded as a special feature of tumor growth \cite{HanahanWein:Cell144:2011:646}.
As the consequence of the transformation of normal cells into cancer cells the immune systems 
reveal different response mechanisms which are described in more detail in 
\cite{Dunn:NatImmun3:2002:991,Kim:Immunology:121:2007:1}. Firstly, the nascent transformed cells 
have to be identified. Candidates for the detection of tumor cells are the components of the innate 
immune system known as Natural Killer cells (NK), Natural Killer T-cells (NKT) and 
so called $\gamma \delta$~T-cells. In case the tumor cells have been recognized the killer cells 
produce the cytokine Interferon-$\gamma$ (IFN-$\gamma$) as an important immunologic regulator 
\cite{Schroder:JLeukBiol:2004:1632004}. Moreover, IFN-$\gamma$ can cause the death of the tumor cell 
directly via apoptotic mechanisms \cite{Wall:ClinCancerRes9:2003:2487}.
The released IFN-$\gamma$ leads to a stimulation of both the innate (activation
of macrophages and presentation of antigens by dendrite cells (DC)) and the
adaptive immune response (generation of antigen-specific B- and T-lymphocytes).
Eventually, the lymphocytes (CD8-positive T cells) migrate to the tumor site,
detect the tumor cells and initiate a powerful immune reaction which may end up
in the destruction of the tumor tissue. The complete suppression of the cancer by the 
immune system is only one scenario. Likewise an imprinting of the tumor cells by their 
immunologic environment can occur during the tumor-immune cells reaction. So a selective 
pressure is exerted on the tumor which favors the creation of tumor cell clones that offer a 
low or even a non-immunogenic behavior. The very different response reflects the paradox role of 
the immune system of cancer promotion due to a sculpting of the immunogenic phenotype of the tumor.
The numerous genetic alterations of the cancer cells during the sculpting process can be regarded as 
a sequence of stochastic events. Therefore, the modeling of the situation in a mathematical model should 
include both deterministic and stochastic parts. In addition the hypothesis of immunoediting suggest the 
occurrence of a phase with metastable states. Within this phase the tumor will neither grow to
its final size nor be eliminated by the immune system. Because the tumor is under immunogenic control such a 
state can be regarded as tumor dormancy. As argued in \cite{Schreiber:Science331:2011:1565} the period 
of this dormant state could be even of the life-time of the host. 
Despite of the short extract of possible effects one realizes that the immune system is a
complex network where a variety of distinct cell types are involved with coordinated functions. 
An essential ingredient is that nearly all different cell types carry more than one functions. So the 
Natural Killer cells are able to release Interferon-$\gamma$ and have simultaneously the ability to 
recognize and eliminate cancer cells. A further example is the immunomodulating agent IFN-$\gamma$ 
which can on the one hand promote the proliferation of lymphocytes and on the other hand can directly 
effect the life of a cancer cell. 

Due to this diversity of cells and their functions and the fact 
that the interplay between tumor and immune cells is far from being understood completely the
development of a mathematical model is necessary. Although one cannot expect that such models 
cover all the underlying biological aspects. Especially a very detailed description of the 
tumor-immune interaction seems not to be realistic. Otherwise such a coarsened model should 
include the main features of the immune system, namely detection, stimulation and elimination of tumor 
cells. Our approach simulates the different functions by introducing two kinds of immune cells named 
tumor detecting cells (TDC) and effector cells. The detecting cells are able to recognize the malignant 
cancer cells and additionally they stimulate the production of effector cells. The last ones have the ability 
to kill tumor cells. Insofar we map the three functions of the immune system onto two artificial cell types, the 
detecting cells and the effector cells. This mapping of the main functions of the immune cells 
allows us to construct a mathematical model the details of which are discussed in the following section.  

\section{Model}

As discussed in the previous section, the immune system of the human
body comprises various components which interact mutually. Moreover, the 
tumor cells are subjected to genetic alterations. Therefore, the tumor system 
can be regarded as being composed of different kinds of cells. In order to present an 
accessible theoretical model of a possible immune reaction against tumor growth 
we refer to the following coarsened description. The tumor system is assumed to 
consist of one single cell type the density of which at time $t$ is denoted as $X(t)$.
Unlike the immune system is realized by two kinds of cells responsible
for detection, stimulation and elimination, respectively.
The elimination process is performed by the effector cells with density $Y(t)$ which are able
to kill the tumor cells. The second immune cell type -the tumor detecting cells
(TDC) designated as $Z(t)$- have the ability to recognize the harmful cancer
cells and in addition stimulate the proliferation of the effector cells.
As basic model the three-state system obeys the following set of deterministic equations 
\begin{align}
\begin{aligned}
\frac{d}{dt}X(t) =& a\,\bigl( X(t)-b\,X^2(t)\bigr) - c\,X(t)\,Y(t) \,,\\
\frac{d}{dt}Y(t) =& e\,Y(t)\,Z(t) - \tilde{\rho} \,Y(t)\,,\\
\frac{d}{dt}Z(t) =& -\tilde{\mu} \,Z(t) \,,
\end{aligned}\label{detmod1}
\end{align}
where the parameters $a,\,b,\,c,\,e,\,\tilde{\rho} ,\,\tilde{\mu} \,>0$ will be discussed now. 
This model incorporate a logistic growth of the cancer cells $X(t)$ with the birth rate $a$.
The undisturbed evolution of the cancer would end when the tumor reaches its
final size -the carrying capacity $b^{-1}$. The effector cells $Y(t)$ can interact with the tumor cells 
and hence the size of the tumor is reduced. The parameter $c$ is a measure for the strength of the 
tumor-effector cell reaction. As suggested in the previous section the effector cells with the ability to 
kill the cancer cells do not exist without an external stimulus. The production of the effector cells 
will be mediated by the TDC with density $Z(t)$. The term $e\,Y(t)\,Z(t)$ in Eq.~\eqref{detmod1} describes 
the initiation of effector cells due to the TDC. The parameter $e$ is the production rate. Because the immune 
system can exert their influence only for a limited period, we have introduced the terms 
$-\tilde{\rho}\,Y(t)$ and $-\tilde{\mu}\,Z(t)$ in Eqs.~\eqref{detmod1}. They reflect the finite lifetime 
$\tilde{\rho}^{-1}$ and $\tilde{\mu}^{-1}$ of the effector cells and the TDC, respectively. As visible from 
Eq.~\eqref{detmod1} an elimination of the tumor is not possible within this approach because a 
release-term for the tumor-recognizing cells is not taken into account and thus effector cells are 
not produced. The three-state model for $(X(t), Y(t), Z(t))$ in Eq.~\eqref{detmod1} offers two stationary 
states $(0,0,0)$ and $(b^{-1},0,0)$ where the tumor-free state $X_s = 0$ is never stable. Instead of 
that the state $X_s= X(t\to \infty )=b^{-1}$ with a finite tumor population is realized. Eq.~\eqref{detmod1} 
predicts that the tumor will always reach its final size determined by the carrying capacity. 
As discussed in Sec. II the tumor-immune interaction is subjected to numerous stochastic events. In the following 
we will demonstrate that random forces are able to create a birth term for the TDC $Z(t)$. As the consequence the 
behavior of the system is changed drastically. To reduce the number of parameters let us introduce dimensionless 
variables according to     
\begin{gather}
\begin{gathered}
x=b\,X\,,\,y=\frac{c}{a}\,Y\,,\,z=\frac{e}{a}\,Z\,,\,\rho
=\frac{\tilde{\rho}}{a}\,,\,\mu =\frac{\tilde{\mu}}{a}\,,\,\bar{t}=a\,t \,.
\end{gathered}\label{scaling}
\end{gather}
In terms of these quantities and under introducing random forces $\eta_i(t)$ the deterministic set of 
Eq.~\eqref{detmod1} is changed to the stochastic differential equations
\be
\frac{d}{dt}x _i(t)= \psi_i[\mathbf{x}(t)]+\Omega_{ij}[\mathbf{x}(t)]\,\eta_j(t)\,.
\label{randomsys}
\ee
Here for simplicity of notation the dimensionless time variable $\bar{t}$ is replaced by $t$ and summation 
over double indices is understood. 
Eq.~\eqref{randomsys} describes the noisy tumor-immune interaction. The vector $\boldsymbol\psi$ and the 
matrix $\boldsymbol\Omega$ are defined by 
\be
\boldsymbol\psi =\begin{pmatrix}
				x-x^2 - x\,y \\
				y\,z -\rho \,y \\
				-\mu \,z
				\end{pmatrix} \qquad ,\qquad
\boldsymbol\Omega =\begin{pmatrix}
						z & 0 & 0 \\
						0 & y & 0 \\
						z & 0 & x
					\end{pmatrix} \,.
\label{psiOmega}
\ee
Further, we have introduced the vector \mbox{$\mathbf{x}=(x,y,z)$} and the vector of the 
stochastic force \mbox{$\boldsymbol\eta =(\eta_x,\eta_y,\eta_z)$}, i.e. the
noise $\eta_i$ is associated with the cell type $x_i$. Eq.~\eqref{randomsys} and 
Eq.~\eqref{psiOmega} include the obvious possibility that the tumor cells $x$ are coupled to 
the random force $\eta_z$ originated in the TDC subsystem. This coupling term appears 
in the equation of motion of the TDC $z$, too. Because the tumor itself is thought to be a source 
of stochastic influences. So the couplings supposed between $z$-cells and the noise force $\eta_x$ 
stemming from the tumor cells. Such a coupling term occurs in the evolution equation of the cancer
cells $x$ as well as in that one of the TDC $z$. The special form of the couplings was chosen to emphasize 
the importance of recognizing the tumor cells and the according stochastic events.
The noisy properties are expressed by the following relations
\begin{align}
\begin{aligned}
\langle \eta_k(t)\rangle =& 0 \,, \\
\chi _{kl}(t,t') =& \langle \eta_k(t)\,\eta_l(t')\rangle 
 = \frac{D_{kl}}{\tau _{kl}}\,\exp\left[-\frac{\mid t-t'\mid}{\tau
_{kl}}\right] \\
 &\xrightarrow[]{\tau _{kl}\rightarrow 0} 2\,D_{kl}\,\delta (t-t') \,.
\label{noise}
\end{aligned}
\end{align}
The components $\eta_k(t)$ have a zero mean. In the limit that the correlation time tends to zero, 
$\tau_{kl} \to 0$, the usual white noise properties are recovered. The correlation strength and
correlation time matrices $\mathbf{D}$ and $\boldsymbol\tau$, respectively, are
assumed to take the forms
\be
 D_{kl}=\begin{pmatrix}
						D_x & S & R \\
						S & D_y & P \\
						R & P & D_z
					\end{pmatrix} \quad ,\quad
\tau_{kl} =\begin{pmatrix}
						\tau & 0 & 0 \\
						0 & \tau & 0 \\
						0 & 0 & \tau
					\end{pmatrix} \quad ,\quad
D_x,\,D_y,\,D_z,\,S,\,R,\,P,\,\tau ,\,>0 \quad.
\label{DandTau}
\ee
The matrix of the correlation time $\tau_{kl}$ reveals that all auto-correlations are
characterized by the finite correlation time $\tau$ whereas the cross-correlation
functions with strengths $R, S$ and $P$ offers white noise properties with the 
$\delta$-function according to Eq.~\eqref{noise}.

\section{Probability distribution}

In this section we derive the probability distribution $P(\mathbf{x},t)$ which is related to the 
set of stochastic equations determined by Eqs.~\eqref{randomsys}-\eqref{DandTau}. Following 
\cite{Gardiner:HandbookStochMeth:Book:1990,Kampen:StochProcPhysChem:Book:1981} we define 
\be
P(\mathbf{x},t)=\bigl\langle \delta \,[\mathbf{x}(t)-\mathbf{x}] \bigr\rangle \,.
\label{probintro}
\ee
Here the $<...>$ means the average over all realizations of the
stochastic process. The vector $\mathbf{x}(t)$ represents the stochastic process
whereas the $\mathbf{x}$ are the possible realizations of the process at time
$t$. Due to the colored noise the corresponding Fokker-Planck equation
can be obtained only approximatively in lowest order of the correlation time.
The time evolution of Eq.~\eqref{probintro} can be written in the form
\be
\frac{\partial}{\partial t}P(\mathbf{x},t)=\mathcal{L}\,P(\mathbf{x},t)\,.
\label{singleevent}
\ee
In deriving this expression we have used the time evolution of $\mathbf{x}(t)$
according to Eq.~\eqref{randomsys}, the Novikov theorem \cite{Novikov:SPJ:20:p1290:1965}
and the correlation function in Eq.~\eqref{noise} with $\tau_{kl}$ and $D_{kl}$
presented in Eq.~\eqref{DandTau}.
The form of the operator $\mathcal{L}$ is given in a correlation time and
cumulant expansion \cite{Fox:JoMP:18:p2331:1977, PhyssicaA:Garrido:1982:479,
Dekker:PLA:90:p26:1982} by
\begin{align}
\begin{aligned}
\mathcal{L}(\mathbf{x})= &-\frac{\partial}{\partial x _i}\psi
_i(\mathbf{x})+D_{kl}\,\frac{\partial}{\partial x _i}\Omega
_{ik}(\mathbf{x})\frac{\partial}{\partial x _n} \Biggl\{ \Omega _{nl}(\mathbf{x})-\tau_{kl} \,M_{nl}(\mathbf{x}) \Biggr.
\\
&+\Biggl. D_{mr}\,\tau_{kl}\,\left[ K_{nlm}(\mathbf{x})\frac{\partial}{\partial x
_s}\Omega _{sr}(\mathbf{x}) + \frac{\tau_{kl}}{\tau_{kl}+\tau_{mr}}\Omega
_{nm}(\mathbf{x})\frac{\partial}{\partial x _s}K_{slr}(\mathbf{x})\right]
\Biggr\} \,,
\label{Lnull}
\end{aligned}
\end{align}
with
\begin{align}
\begin{aligned}
M_{nl} = \psi _r\frac{\partial \Omega _{nl}}{\partial x _r}-\Omega
_{rl}\frac{\partial \psi _n}{\partial x _r}\,,\quad 
K_{nlk} = \Omega _{rk}\frac{\partial \Omega _{nl}}{\partial x
_r}-\frac{\partial \Omega _{nk}}{\partial x _r}\Omega _{rl}\,.
\label{MKQ}
\end{aligned}
\end{align}
The single probability distribution is determined by the operator $\mathcal{L}$ in Eq.~\eqref{Lnull}. 
Notice that the representation is valid for sufficiently large times scale compared with the
correlation times $\tau_{kl}$ when transient terms are negligible.
Eqs.~\eqref{singleevent}-\eqref{MKQ} enable to find the equation of motion for
the expectation values $\left\langle x_j(t) \right\rangle$. It follows
\begin{align}
\begin{aligned}
\frac{d}{dt}\left\langle x _j(t)\right\rangle =& \left\langle \psi
_j\right\rangle +D_{kl}\,\left\langle \frac{\partial \Omega _{jk}}{\partial x
_n}\bigl( \Omega _{nl}-\tau_{kl}\,M_{nl} \bigr) \right\rangle \\
&-D_{kl}\,D_{mr}\,\tau_{kl} \,\Biggl\{ \left\langle \frac{\partial}{\partial x
_s} \left( \frac{\partial \Omega _{jk}}{\partial x _n}K_{nlm}\right) \Omega
_{sr}\right\rangle \Biggr. \\
&+\Biggl. \frac{\tau_{kl}}{\tau_{kl}+\tau_{mr}}\,\left\langle \frac{\partial}{\partial x
_s} \left( \frac{\partial \Omega _{jk}}{\partial x _n}\Omega _{nm}\right)
K_{slr} \right\rangle \Biggr\}\,.
\label{EWxyz}
\end{aligned}
\end{align}
Remark that in the limiting case of white noise all terms including $\tau_{kl}$ vanish.
Further we want to point out that the expression in Eq.~\eqref{EWxyz} contains
quadratic terms like $\langle x_i\,x_j\rangle$ due to the nonlinear system in
Eq.~\eqref{randomsys}. In the same manner as before one can derive a higher order joint probability 
distribution, see \cite{Hernandez-Machado:ZFPBM:52:p335:1983}. Following this procedure we 
get a whole hierarchy of evolution equations. Instead of that let us make the simplest approximation 
\mbox{$\langle x_i\,x_j\rangle = \langle x_i\rangle \langle x_j\rangle$}. Under this approximation 
Eq.~\eqref{EWxyz} and by applying Eq.~\eqref{MKQ} the equation of
motion for the mean values can be rewritten as
\begin{align}
\begin{aligned}
\frac{d}{dt}\langle x(t)\rangle =& [1+R(1-D_x \tau)+\frac{1}{2} D_x
D_z\tau]\,\langle x(t)\rangle - \langle x(t)\rangle^2 - \langle x(t)\rangle
\,\langle y(t)\rangle +[D_x(1+R \tau)]\,\langle z(t)\rangle\,,\\
\frac{d}{dt}\langle y(t)\rangle =& \langle y(t)\rangle\,\langle z(t)\rangle -
(\rho -D_y)\,\langle y(t)\rangle \,,\\
\frac{d}{dt}\langle z(t)\rangle =& [R(1-D_x \tau)+D_x D_z \tau]\,\langle
x(t)\rangle -[\mu -(R(1+D_x \tau )+D_x(1+\frac{1}{2} D_z \tau))]\,\langle
z(t)\rangle \,.
\end{aligned}\label{modfinal}
\end{align}
As can be seen from Eq.~\eqref{modfinal} the random process referring to the correlation
strength and correlation time presented in Eq.~\eqref{DandTau} 
influences the dynamical system in a significant manner. The behavior is illustrated 
in Fig.~\ref{figmodel}.
\bef
\includegraphics[width=8cm]{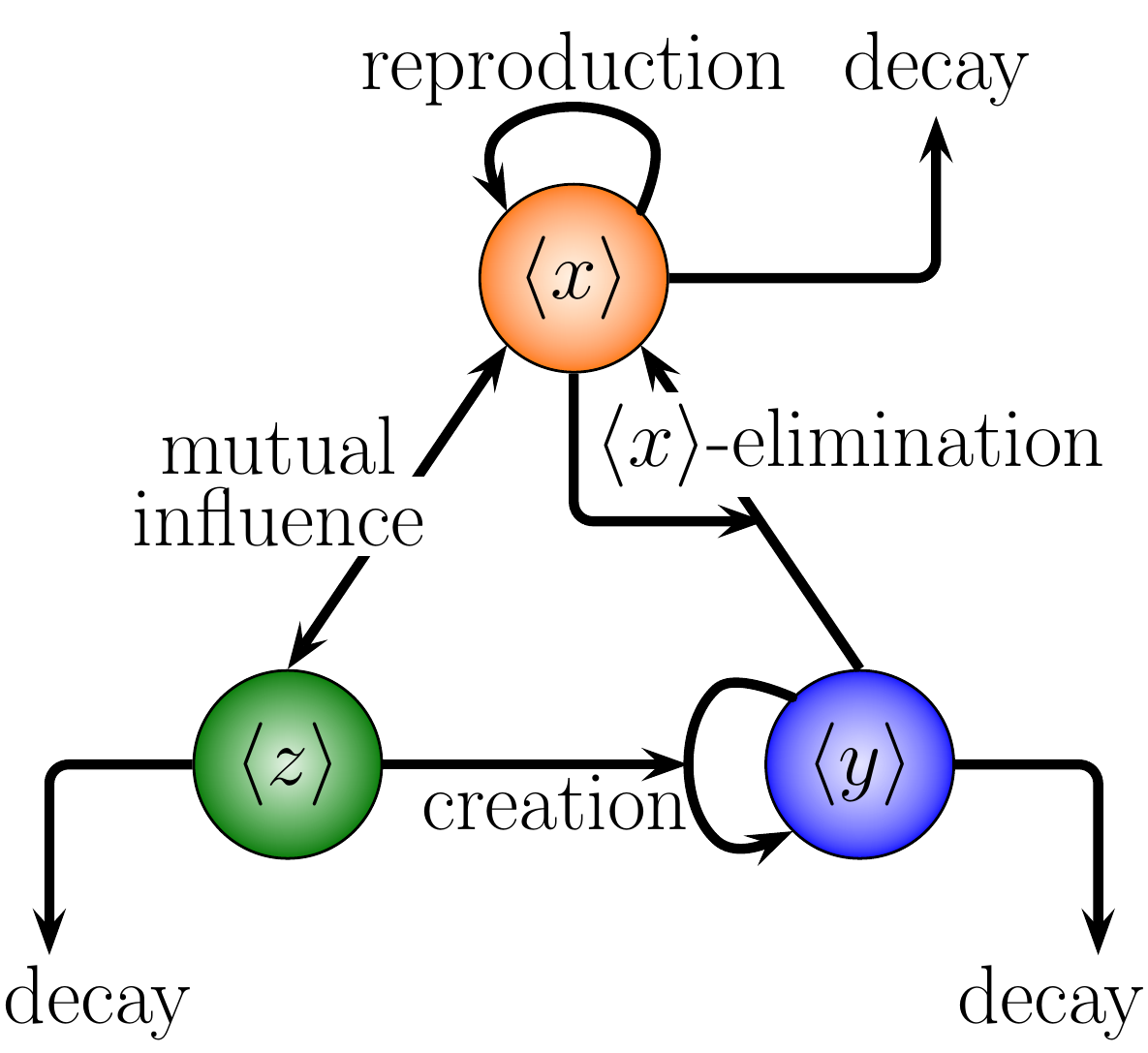}
	\caption{(Color online) Schematic illustration of the model presented in
Eq.~\eqref{modfinal}.}
	\label{figmodel}
\ef
Let us compare the results of the stochastic approach with the deterministic model. The 
birth rate of the tumor cells $\langle x\rangle$ are affected by the noise correlation strengths 
$D_x$, $D_z$ associated with the tumor cells and the tumor detecting cells as well as 
their cross-correlation $R$. Likewise the decay rate $\rho$ in the equation for the effector cells
$\langle y\rangle$ is reduced by the noise strength $D_y$, i.e. by the noise
related to the effector cell subsystem itself. As a new nontrivial result we find noise induced 
terms in the evolution equation Eq.~\eqref{modfinal}. So there appears a term \mbox{$\propto \langle x(t)\rangle$} 
in the equation for the $z$-cells which are able to recognize cancer cells. In the same manner 
the generating term \mbox{$\propto \langle z(t)\rangle$} arises in the equation for the tumor
cells. These terms are originated exclusively due to the randomness in
Eqs.~\eqref{randomsys} and \eqref{psiOmega}. In this context the most important parameter is played by 
the cross correlation strength $R$ of the correlation function
$\chi_{xz}=\chi_{zx}$, i.e. the correlation between the noise sources inherent in the tumor cells 
and the tumor detecting cells. Notice that in the noise-free case such an interlink between these two cells 
is missing, compare again Fig.~\ref{figmodel}.
Moreover, the death rate $\mu$ is altered due to the stochastic process. Based on the implementation of 
noisy forces the resulting Eq.~\eqref{modfinal} differs from the deterministic equation twice. 
(i) Firstly, the birth and death rates as well are altered due to stochastic parameters  
such as the correlation time $\tau$ and the correlation strengths $D_x$, $D_y$, $D_z$ and $R$ defined in 
Eq.~\eqref{DandTau}. Although the cross-correlations $S$ and $P$ were included in
Eqs.~\eqref{randomsys} and \eqref{psiOmega}, they do not appear in the final expression Eq.~\eqref{modfinal}. 
This fact is related to the special kind of multiplicative noise of our model. 
(ii) Secondly, two new terms exist in Eq.~\eqref{modfinal}. The
origin of both can be solely ascribed to stochastic sources. Regarding the
evolution of the tumor detecting cells $\langle z\rangle$ the new term disappears in case $R=0$ and $\tau =0$. 
So both parameters $R$ -the cross-correlation strength between $\eta_x$ and $\eta_z$ and $\tau$ -the 
correlation time of the auto-correlation functions \mbox{$\langle \eta_i(t) \eta_i(t') \rangle$}-
are of significant relevance. In the subsequent section the analysis is focused in particular on 
both parameters.

\section{Results and discussion}

As remarked in the previous section the parameters $R$ and $\tau$ inhere a special meaning in discussing 
the set of Eq.~\eqref{modfinal}. Before we proceed with the stability analysis and the results we want to
estimate the model parameters. The starting point is the deterministic Eq.~\eqref{detmod1}.
One finds different values for the intrinsic tumor growth rate $a$: $0.18~\rm{day}^{-1}$ 
\cite{Kuznetsov:BullMathBiol:56:1994} and $0.51~\rm{day}^{-1}$ \cite{DePillis:CancerRes:65:7950:2005}. Our own 
study leads to $0.57~\rm{day}^{-1}$ \cite{BoseTrimper:PRE:2009:051903}. The first two values are based on mouse 
models while the latter one was obtained by means of {\em in vitro} cultivation of tumor cells. The growth 
rate is insofar of importance as it determines the time scale of the dynamics, see Eq.~\eqref{scaling} 
($t~[\rm{in\, days}]=\bar{t}/a$). Here we use $a=0.5~\rm{day}^{-1}$. Thus, $\bar{t}=1$ is tantamount to 
$t=2~\rm{days}$. An estimation of the carrying capacity is $b^{-1}=10^9~\rm{cells}$ 
\cite{Kuznetsov:BullMathBiol:56:1994,DePillis:CancerRes:65:7950:2005}. Further, the reaction rates take 
approximately $c=10^{-7}~\rm{cell}^{-1}~\rm{day}^{-1}=e$ 
\cite{Kuznetsov:BullMathBiol:56:1994,DePillis:CancerRes:65:7950:2005,
KirschnerPanetta:JMathBiol:1998,Rodriguez-P:MathMedBiol:24:2007}. An estimation for the decay rates in 
Eq.~\eqref{detmod1} is given by $\tilde{\rho}=3\times 10^{-2}~\rm{day}^{-1}$ and 
$\tilde{\mu}=10~\rm{day}^{-1}$ 
\cite{DePillis:CancerRes:65:7950:2005,KirschnerPanetta:JMathBiol:1998}. In relating our results to real 
units one should take into account the scaling properties Eq.~\eqref{scaling}. All the results are collected 
at the end of this section in Tab.~\ref{Table1}. For the subsequent analysis it is more convenient to 
use dimensionless quantities. The both most relevant parameters of stochastic forces are the 
auto-correlation time $\tau$ and the cross-correlation strength $R$. Both quantities $R$ and $\tau$ will be 
altered within the interval $[0,5]$. The remaining parameters are assumed to be fixed, i.e. 
$D_x=2.1$, $D_z=1.2$ and $D_y=0.01$. The values for $D_x$ and $D_z$ are chosen arbitrarily, whereas the 
value for $D_y$ is suggested to be smaller than $\rho=\tilde{\rho}/a=0.06$ in order to guarantee a sufficient  
stability of the differential equation system Eq.~\eqref{modfinal}, cf. the term $\propto \langle
y\rangle$.
Moreover, since we consider cell populations the solutions of
Eq.~\eqref{detmod1} should yield positive values for the cell numbers. So values of $R$ and $\tau$ 
are excluded when they induce negative values for the cell populations.

Now we perform the stability analysis according to the tumor-immune cells
reaction system satisfying Eq.~\eqref{modfinal}. We note that the numerical bifurcation analysis is performed by means 
of the program \cite{xppaut541} which contains the bifurcation tool \cite{autobif}. This set of equations exhibits three different
equilibria, i.e. the tumor-free $E_1=(0,0,0)$, and two non-tumor-free states
designated as $E_2$ and $E_3$. The last ones are given by lengthy expressions in terms of the model 
parameters. Only one of the three equilibria is stable simultaneously. It is also possible
that the total system becomes unstable as discussed below.
The solution of Eq.~\eqref{modfinal} depends strongly on the correlation time
$\tau$ and the cross-correlation strength $R$. Concerning $\tau$ we find
three different regions (labeled as I-III) where the solution of
Eq.~\eqref{modfinal} has different properties. 
The threshold values referring to our specific numerical values of the remaining
model parameters are $\tau_{c_1}=0.636$ and $\tau_{c_2}=4.016$.
We proceed by considering these three regions determined by the correlation time
$\tau \in [0,5]$. As fixed initial values for the tumor and the tumor detecting cells, 
respectively, we choose $\langle x(t=0)\rangle =10^{-6}$ and $\langle z(t=0)\rangle =0$ . This reflects a situation 
where the tumor is small and tumor detecting cells are not present. In our case this equals an initial tumor cell number of 
$10^3~\rm{cells}$ which is clinically not detectable (early stages of tumor evolution).
\bef
\centering
\subfigure[~$\tau =0.3 \to$ region~I]{
				\label{figbifber1}
				\includegraphics[width=0.46\linewidth
]{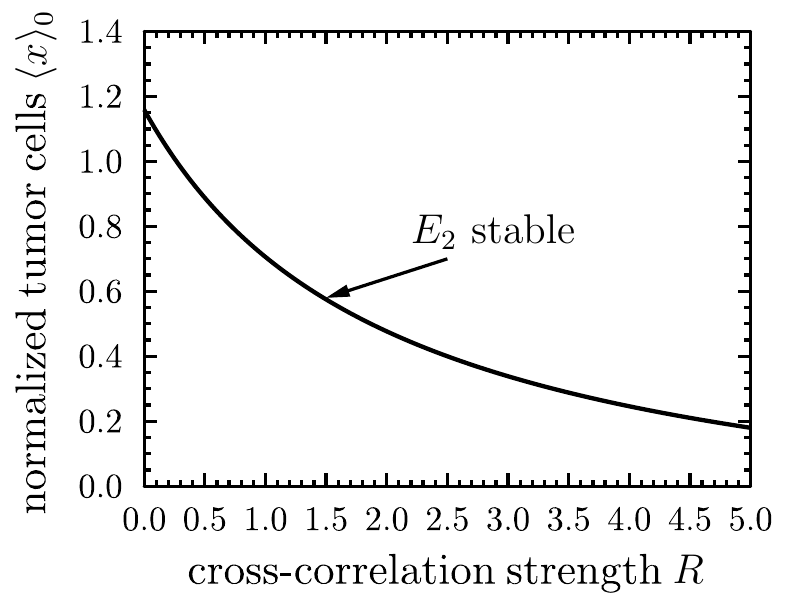}}
\hspace{0.05\linewidth}
\subfigure[~$\tau =2.0 \to$ region~II]{
				\label{figbifber2}
				\includegraphics[width=0.46\linewidth
]{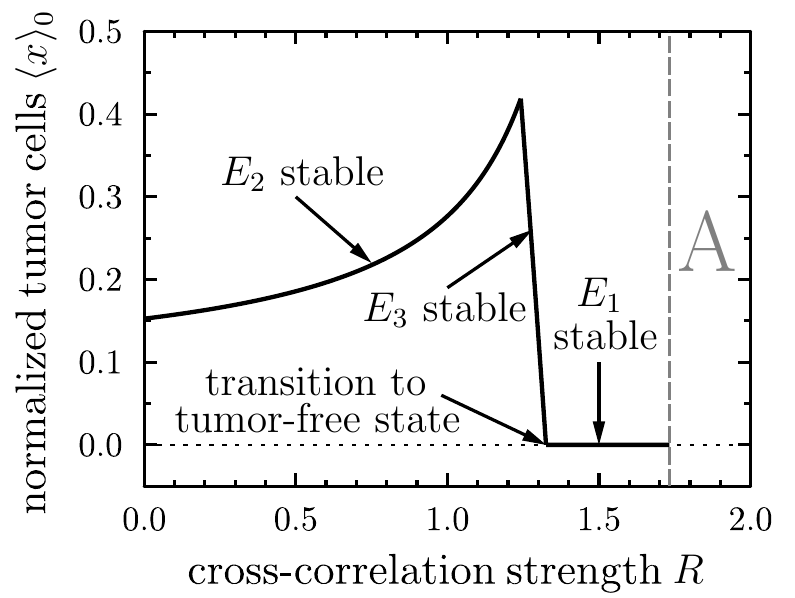}}\\[1em]
\subfigure[~$\tau =4.5 \to$ region~III]{
				\label{figbifber3}
				\includegraphics[width=0.46\linewidth
]{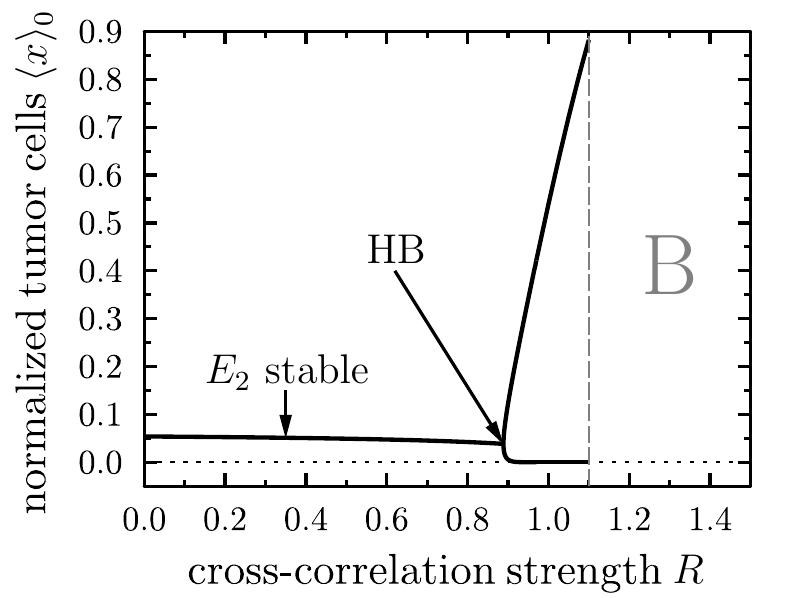}}
\hspace{0.05\linewidth}				
\subfigure[~$\tau =4.5 \to$ region~III]{
				\label{figsolber3b}
				 \includegraphics[width=0.46\linewidth
]{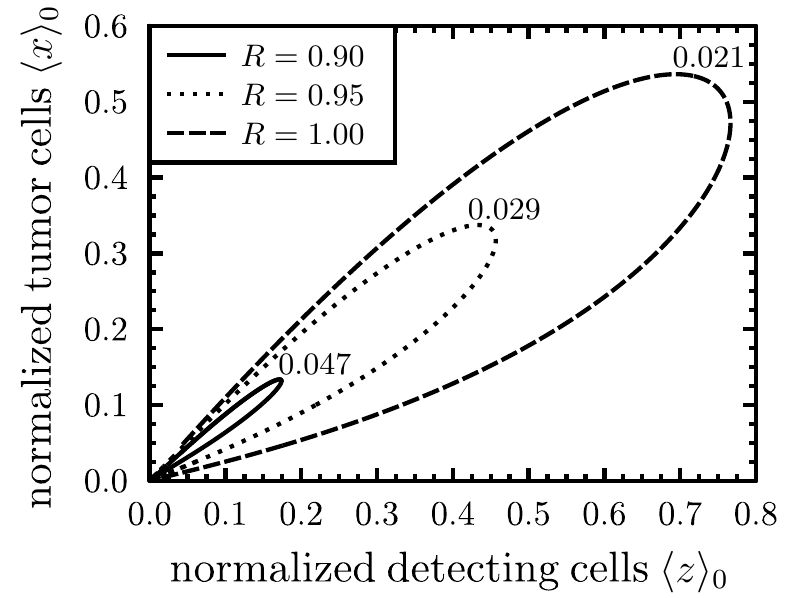}}
\caption{Behavior of the solution representing the regions~I-III mentioned in the text.
The parameters take $\rho =0.06$, $\mu =20$, $D_x=2.1$, $D_y=0.01$ and $D_z=1.2$.
(a)-(c):~Bifurcation diagrams. (d)~Limit cycles in the
$\langle x\rangle -\langle z\rangle -$pahse space.}
\label{figresult1}
\ef
Tumor detecting cells should be only generated due to stochastic
influences and did not exist a priori. For the initial state of the effector cell
population we have to choose a non-zero value of $\langle y(t=0)\rangle >0$.
Otherwise the solution will differ from the predictions based on the stability
analysis which is due to the structure of the differential equation system in
Eq.~\eqref{modfinal}. In some regions this value can be very small (almost zero, e.g. $\sim 10^{-20}$)
whereas in other areas corresponding to the parameters $R$ and $\tau$ the
stability of the solution depends on $\langle y(t=0)\rangle$ significantly .
In what follows these distinct cases will also be discussed while we want to
restrict the possible values for $\langle y(t=0)\rangle$ to the left-open
interval $(0,10]$.

{\em Region~I} ($0\leq \tau \leq \tau_{c_1}=0.636$): Within this range a stable tumor-free
state is missing for $R \in [0,5]$. All three fixed points exist in this region. The
solution tends either to the steady states $E_2$ or $E_3$ one of which is
asymptotically stable which depends on the cross-correlation strength $R$. 
For instance in case of $\tau =0.3$ the equilibrium value of the tumor cell population, 
designated as $\langle x\rangle_0$ as a function of $R$, is depicted in
Fig.~\ref{figbifber1}.
As is visible for all $0\leq R\leq 5$ the solution will always reach the fixed
point $E_2$. Further, the equilibrium value $\langle x\rangle_0$ decreases with
increasing $R$. In region~I the initial value of the effector cells can take
arbitrary positive values in $(0,10]$ without changing the solution of
Eq.~\eqref{modfinal}.
An exemplary dynamical solution is illustrated in Fig.~\ref{figsolber1}.

{\em Region~II} ($\tau_{c_1}=0.636 < \tau <\tau_{c_2}=4.016$): In this area the behavior
is changed and one observes diagrams like that one shown in Fig.~\ref{figbifber2} for $\tau =2.0$. 
The three fixed points $E_{1,2,3}$ survive for $R \in [0,5]$, but the stability is changed.
If we start at $R=0$ and increase the cross-correlation strength $R$ the
behavior of the solution traverses four different regions. For $0\leq R< 1.242$ the steady state 
$E_2$ is stable. At $R=1.242$ a transcritical bifurcation occurs where $E_2$ is not stable anymore.
The fixed point $E_3$ becomes stable but only within the interval $1.242\leq R<1.325$. At 
$R=1.325$ another transcritical bifurcation occurs namely the transition to
the tumor-free state $E_1$ which becomes stable while $E_3$ loses its stability. 
Biologically such a transition is of great relevance because it manifests that the 
immune system is able to eliminate a growing tumor provided the tumor-immune
cells reaction is assisted by a cross-correlation between stochastic events occurring 
in the tumor and in the tumor detecting cells subsystem. Notice that the sector A in Fig.~\ref{figbifber2} 
has to be excluded because the eigenvalues of $E_1=(0,0,0)$ develop an imaginary part indicating the solution 
tends to $E_1$ on a stable spiral. However, during the evolution towards the
equilibrium value ($\langle x(t)\rangle \to \langle x\rangle_0$, $t\to \infty$)
the tumor cell population $\langle x(t)\rangle$ takes negative values. This
happens for $R>1.733$ which is indicated by the sector A in
Fig.~\ref{figbifber2}.
In the area $R\leq 1.733$ there are no restrictions on the initial value
for the effector cells $0< \langle y(0)\rangle \leq 10$.
In Fig.~\ref{figsolber2} the time evolution of the tumor cell number $\langle x(t)\rangle$
is shown for different values of the cross-correlation strength $R$.
Summarizing the result we observe in region~II the occurrence of tumor escape
as well as the possibility of tumor elimination depending on the value of the
cross-correlation strength $R$.
\bef
\centering
\subfigure[~$\tau =0.3 \to$ region~I]{
				\label{figsolber1}
				\includegraphics[width=0.46\linewidth
]{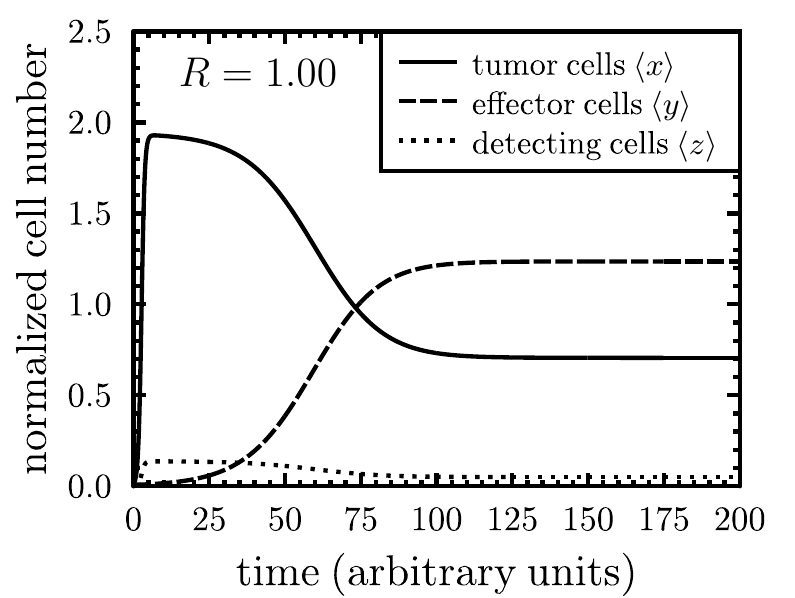}}
\hspace{0.05\linewidth}
\subfigure[~$\tau =2.0 \to$ region~II]{
				\label{figsolber2}
				\includegraphics[width=0.46\linewidth
]{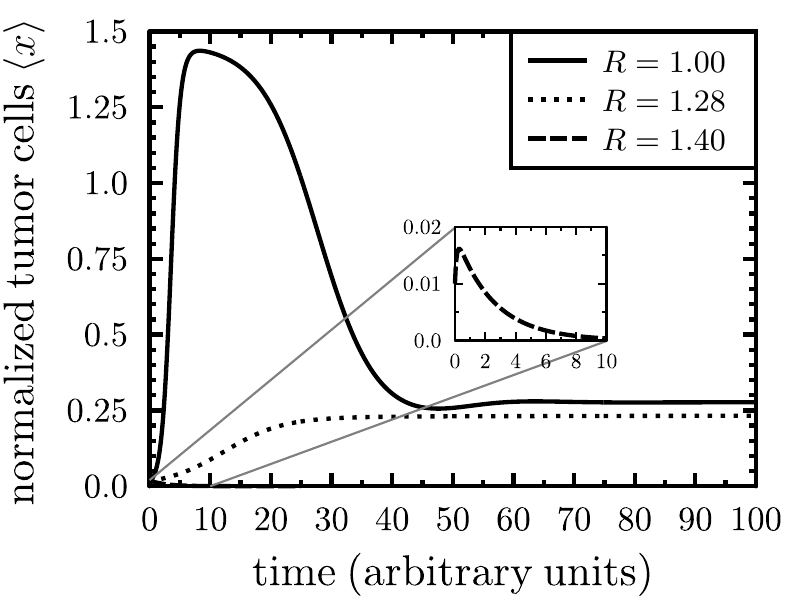}}\\[1em]
\subfigure[~$\tau =4.5 \to$ region~III]{
				\label{figsolber3a}
				\includegraphics[width=0.46\linewidth
]{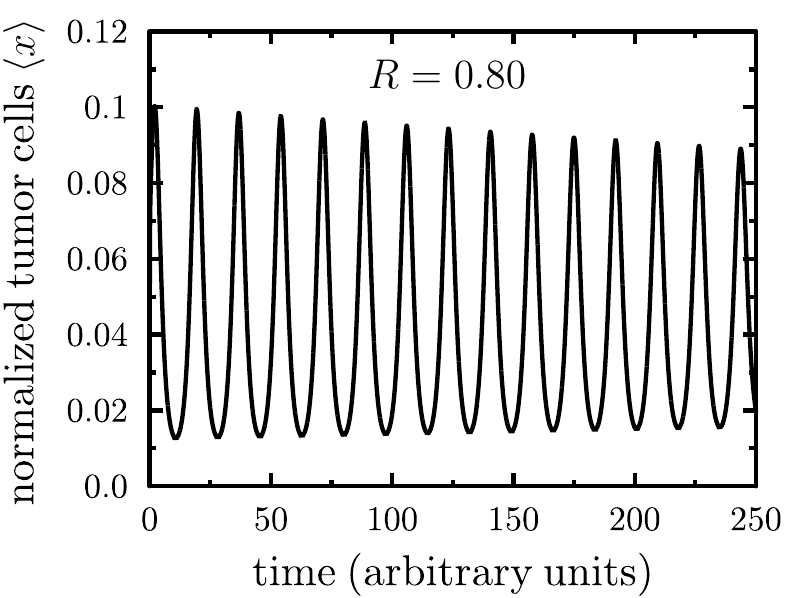}}
\hspace{0.05\linewidth}				
\subfigure[~$\tau =4.5 \to$ region~III]{
				\label{figsolber3c}
				 \includegraphics[width=0.46\linewidth
]{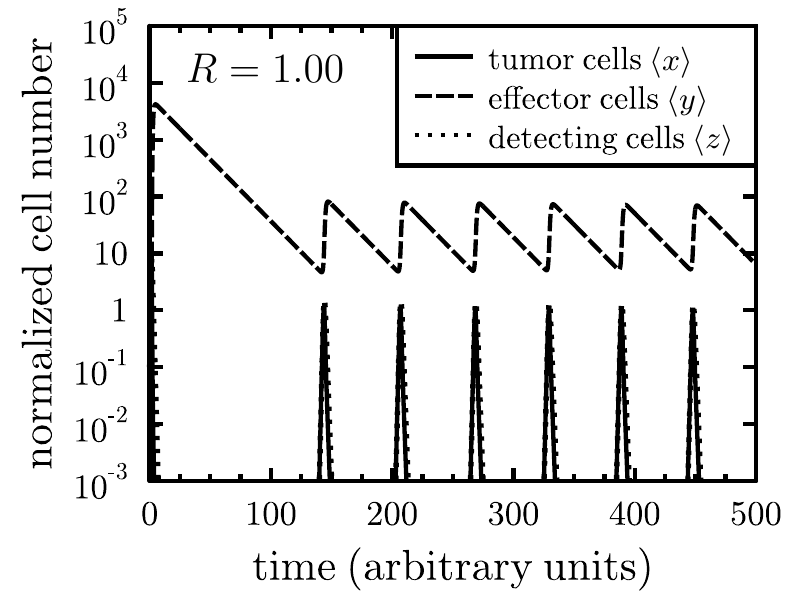}}
\caption{Exemplary dynamic solutions according to the regions~I-III mentioned in the text.
The parameters take $\rho =0.06$, $\mu =20$, $D_x=2.1$, $D_y=0.01$ and $D_z=1.2$.}
\label{figresult2}
\ef

{\em Region~III} ($\tau_{c_2}=4.016 \leq \tau \leq 5$): In this parameter range 
one observes a new behavior determined by the cross-correlation strength $R$ and the
initial value of the effector cells $\langle y(0)\rangle$. For the following discussion 
we refer to Fig.~\ref{figbifber3}. Starting from $R=0$ and increasing this parameter 
the solution of Eq.~\eqref{modfinal} tends to the stable fixed point $E_2$. A related solution 
is represented in Fig.~\ref{figsolber3a}, where it needs a rather long time until $E_2$ is reached, 
about $t\sim 19000$. The fixed point $E_2$ is realized on a stable spiral within the interval 
$0\leq R<0.888$. The smaller the cross-correlation strength $R$ is the shorter is the time scale to 
reach $E_2$, for instance we need $t=1000$ in case $R=0.01$. 
When the critical value $R=0.888$ is exceeded a periodic limit cycle evolves related to the occurrence of 
a Hopf bifurcation. In Fig.~\ref{figbifber3} the minimal and the maximal numbers of tumor cells 
are plotted within such a limit cycle. The numerical values range below and
above the former stable equilibrium $E_2$ which becomes now unstable. After the Hopf bifurcation the 
steady states $E_1$ and $E_3$ are no longer detectable. Further, Fig.~\ref{figbifber3} 
reveals that the the parameter range is limited in which such stable periodic oscillations emerge.
The dashed line represents the boundary to sector B, where the total system bifurcates into an unstable
state and the dynamical system is uncontrollable anymore. Thus the sector B will be excluded as a domain 
of accessible solutions within our tumor-immune model.
Nevertheless, periodic orbits can be observed for $0.888 \leq R\leq 1.101$ and fixed correlation time 
$\tau =4.5$. For varying values of $R$ the periodic solutions are depicted in the two-dimensional 
$\langle x\rangle -\langle z\rangle -$phase space, see Fig.~\ref{figsolber3b}.
The numbers shown above each orbit is the frequency of the oscillations between two maxima. With growing $R$ from $0.9$ to $1.0$ 
the minimal and the maximal cell numbers increase for both $\langle x\rangle$ and $\langle z\rangle$ while  
the frequency decreases, i.e. the period of the oscillation is enlarged. This result is also valid for the 
effector cells $\langle y\rangle$ which are not shown here.
\bef
\includegraphics[width=0.46\linewidth ]{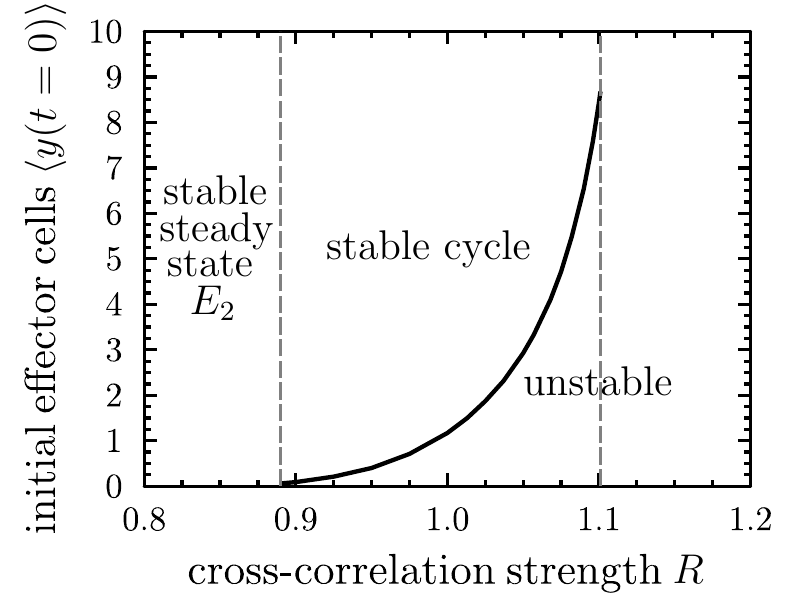}
\caption{Distinction between stable and unstable solutions depending on
$\langle y(0)\rangle$ and $R$. Further description in the text.}
\label{figRvonY0}
\ef
Coming back to the influence of the initial values of the effector cells $\langle y(0)\rangle$. As already mentioned 
they play a decisive role in the regime of periodic limit cycle solutions. 
For very low initial values the system loses its stability and the solution is
not accessible biologically. The curve which separates stable periodic cycles from unstable solutions is
displayed in Fig.~\ref{figRvonY0}. The question arises what does it means for the real tumor-immune cells 
interactions if stable periodic oscillations occur? In that case we argue that an intensive interaction 
between the cancer and the immune cells takes place at the beginning of the tumor growth as well as 
after a long period later. If the nascent transformed cells start to grow up the immune system is able to 
detect this harmful process and it responds. Such an immune attack reduces the tumor size without to 
delete it completely. Notice that our numerical estimation represented in Fig.~\ref{figbifber3} is also 
compatible with an elimination of the tumor cells because the lower branch is partly not distinguishable 
from zero in the interval $0.888\leq R\leq 1.101$. But the tumor starts anew to grow up signalizing 
the latent facility that the tumor evolution goes on. Otherwise, if the cancer growth is continued the 
immune system remains active and consequently it is still able to eliminate a large amount of tumor cells.
So after a certain time one expect that a balance between tumor growth and the response of the immune system 
is evolved. From here one concludes that the tumor is under the control of the immune system and a so
called tumor dormant state emerges. In the same manner the region with $R<0.888$ (before the Hopf bifurcations 
appears) may also be interlinked to the tumor dormant state. In that case the number of tumor cells is low  
compared to other parameter regimes, see Fig.~\ref{figresult1}. For instance the value 
$R=0.6$ yields an equilibrium tumor cell population of $5\%$ of the carrying capacity in Fig.~\ref{figbifber3}.
However, as a result of our computations, the size of the maximal number of tumor cells within one cycle 
can take large values, e.g. for $R=1.1$ we find $\langle x\rangle_{max}=0.88$. The tumor reaches $88\%$ of its
carrying capacity. The time-dependent solution for $R=1.0$ is depicted in Fig.~\ref{figsolber3c} within 
the interval $0\leq t\leq 500$. Eventually a periodic cycle with $\langle x\rangle_{max}=0.54$ will be reached 
after $t\sim 12000$.

At the end of this section we want to convert some dimensionless quantities into
quantities with real units. The results are particularized in Tab.~\ref{Table1}. 

\begin{table}[htb]
    \centering
\begin{tabular}{lcc}
\hline \hline
quantity \rule[0pt]{0.5cm}{0pt}& \rule[0pt]{0.2cm}{0pt}arbitrary units\rule[0pt]{0.2cm}{0pt}
& \rule[0pt]{0.2cm}{0pt}real units\rule[0pt]{0.2cm}{0pt} \\ \hline \hline
time t & $1$ & $2~\rm{days}$ \\
auto-correlation time $\tau$ & $1$ & $2~\rm{days}$ \\
cross-correlation strength $R$ & $1$ & $0.5~\rm{day^{-1}}$ \\
number of tumor cells $\langle x\rangle$ & $1$ & $10^9~\rm{cells}$ \\
number of effector cells $\langle y\rangle$ & $1$ & $5\times 10^6~\rm{cells}$ \\
number of tumor detecting cells $\langle z\rangle$ & $1$ & $5\times 10^6 ~\rm{cells}$ \\
frequency (period) of cycles & $0.029$ & $\approx 0.01~\rm{day}^{-1}$ \\[-1.7ex]
according to Fig.~\ref{figsolber3b} & ($\approx 34.5$) & ($\approx 69~\rm{days}$) \\ \hline \hline
\end{tabular}
\caption{\label{Table1}Comparison of model quantities in arbitrary and real units.
See also Eq.~\eqref{scaling}.}
\end{table}
Not commented yet
is that the strengths in the correlation functions in Eq.~\eqref{noise} carry the unit
$\rm{day}^{-1}$ after conversion to real units. This follows from the correlation function in
real units, i.e. \mbox{$\langle \eta_i \eta_j' \rangle \,a^2\propto (D_{ij}\,a)/(\tau_{ij}/a)$} where
the intrinsic tumor growth rate $a=0.5~\rm{day^{-1}}$ and $D_{ij}$ and $\tau_{ij}$ are given
in arbitrary units. Thus, the strengths occurring in the noise-noise correlation functions in
Eq.~\eqref{noise} have the meaning of a rate.

\section{Conclusions}

We have presented a mathematical model for the tumor-immune cells reactions which is essentially 
supplemented by stochastic forces. The parameters of the noise correlation function have a
great impact on the behavior of the coupled tumor-immune cell interaction, especially on the 
response of the immune system. In particular we have emphasized that the auto-correlation time $\tau$ 
and the cross-correlation strength $R$ are able to control the evolution of the tumor. More precise, 
these two quantities discriminate whether the system tends to tumor suppression, tumor progression
or tumor dormancy. The assistance of an inevitable noisy influences seems to play a crucial role
during cancer genesis and growth in humans. The involved random forces may be
originated within the tumor as well as inside the immune system and can even interact mutually which is 
manifested in the cross-correlation. Our model should be considered as an attempt toward a more detailed 
analysis of tumor-immune systems. But also the model studied elucidates that noise plays an 
decisive role in such systems.    
The model can be refined immediately, e.g. a finite correlation time is attributed to the 
cross-correlation functions, too. In that case the correlation time matrix, Eq.~\eqref{DandTau}, is 
modified and new terms in Eq.\eqref{modfinal} occur. We believe that our approach includes the most 
relevant degrees of freedom. \\

One of us (T.B.) is grateful to the Research Network 'Nanostructured
Materials'\,, which is supported by the Saxony-Anhalt State, Germany.

\clearpage

\bibliography{BibNATIR}
\bibliographystyle{apsrev4-1}
\end{document}